\documentclass[reprint, 
superscriptaddress,
 amsmath,amssymb,
 aps, physrev,
]{revtex4-2}

\usepackage{graphicx}
\usepackage{dcolumn}
\usepackage{bm}
\usepackage{hyperref}

\usepackage{placeins}
\usepackage{color}
\usepackage{xcolor}
\usepackage{lipsum}
\usepackage{tikz}
\usepackage{caption}
\usepackage{subcaption}
\usepackage{amsmath}
\definecolor{darkseagreen}{rgb}{0.56, 0.74, 0.56}
\definecolor{viridian}{rgb}{0.25, 0.51, 0.43}
\definecolor{shamrockgreen}{rgb}{0.0, 0.62, 0.38}
\definecolor{copper}{rgb}{0.83, 0.69, 0.22}
\definecolor{darkbrown}{rgb}{0.4, 0.26, 0.13}
\definecolor{tropicalrainforest}{rgb}{0.0, 0.46, 0.45}
\definecolor{MatLabBlue}{HTML}{0072BD}
\definecolor{MatLabRed}{HTML}{D95319}

\usetikzlibrary{decorations.pathreplacing}

\begin{document}

\preprint{APS/123-QED}

\title{\textbf{Enhanced nanocomposite susceptibility by field-alignment of superparamagnetic particles}
}%

\author{M. Zambach}
\affiliation{DTU Physics, Technical University of Denmark, 2800 Kgs. Lyngby, Denmark}

\author{T. Veile}
\affiliation{DTU Physics, Technical University of Denmark, 2800 Kgs. Lyngby, Denmark}

\author{M. Varón}
\affiliation{DTU Physics, Technical University of Denmark, 2800 Kgs. Lyngby, Denmark}

\author{M. Knaapila}
\affiliation{Department of Physics, Norwegian University of Science and Technology, 7491 Trondheim, Norway}

\author{L. Almásy}
\affiliation{Institute for Energy Security and Environmental Safety, HUN-REN Centre for Energy Research, 1121 Budapest, Hungary}

\author{T. S. Plivelic}
\affiliation{MAXIV Laboratory, Lund University, 224 84 Lund, Sweden}

\author{C. Johansson}
\affiliation{RISE Research Institute of Sweden, Sensor Systems, 417 55 Göteborg, Sweden}

\author{Z. Ouyang}
\affiliation{DTU Electro, Technical University of Denmark, 2800 Kgs. Lyngby, Denmark}

\author{M. Beleggia}
\affiliation{DTU Nanolab, Technical University of Denmark, 2800 Kgs. Lyngby, Denmark}
\affiliation{Department of Physics, University of Modena and Reggio Emilia, 41125 Modena, Italy}

\author{C. Frandsen}\email{Contact author: fraca@fysik.dtu.dk}
\affiliation{DTU Physics, Technical University of Denmark, 2800 Kgs. Lyngby, Denmark}

\date{\today}

\begin{abstract}
Nanocomposites comprised of insulated magnetic single-domain particles are promising candidates for high-frequency, eddy current free, soft magnetic materials, but tend to suffer from low magnetic susceptibility ($<20$). Particle alignment has been proposed to increase nanocomposite susceptibility and reduce magnetic losses but experimental verification has been lacking.
Here, magnetic nanocomposites containing 3-57 vol\% field-aligned 11$\pm$3 nm maghemite ($\gamma$-Fe$_2$O$_3$) particles in a poly-vinyl matrix were investigated for potential use as high-frequency inductor core materials. The particles were aligned by a homogenous static alignment field during nanocomposite drying, fixating the particle orientation. Particle aggregation was disproved by small-angle scattering. The dependence of the alignment field strength and particle concentration on the nanocomposite's susceptibility and hysteresis losses were investigated from DC up to 922 kHz by vibrating sample magnetometry, AC-susceptibility and high-frequency hysteresis measurements.
Interestingly, the nanocomposite susceptibility increased super-linearly with particle fraction due to weak inter-particle interactions. Alignment of the particles increased the nanocomposite susceptibility from 21 to 50 for samples with a particle content of 57 vol\%. Hence, the synergy between particle alignment and interaction allows for a higher than expected susceptibility of nanocomposites. Furthermore, the results show that magnetically aligning particles in a nanocomposite reduces magnetic losses when using well-dispersed single-domain superparamagnetic nanoparticles. The measured nanocomposite susceptibility could be modelled by a combination of directional dependent Debye-models including mean-field interaction effects and partial particle alignment. The resulting susceptibility of 50 is among the highest obtained for nanocomposites, making it a relevant candidate for applications in power electronics. 
\end{abstract}

\maketitle

\section{Introduction}
Soft magnetic materials with high susceptibility (or high permeability) are key for power converters. Up to now, ferrite materials have been optimised for high susceptibility ($\chi>800$) to be operated below 2 MHz. Above this frequency, eddy-currents become problematic and the susceptibility drops rapidly \cite{petrecca2019a,TDK}. Ferrite materials developed for higher frequency operation usually have a lower susceptibility ($\chi\leq40$), can be operated up to 50 MHz, but suffer from eddy-current losses \cite{Fairite}. Magnetic nanocomposites have been proposed as an alternative core materials for inductors in power electronics, due to their potentially lower losses and higher frequency cut-off compared to the typically used ferrite materials \cite{petrecca2019a,he2023a,hurley2018a,Bima_Preprint,hurley2018a,PSMA_yawger2022a}. Theoretical investigations show that composites containing single-domain magnetic nanoparticles in an electrical insulating polymer matrix could have higher susceptibility and lower losses than the ferrites used today \cite{zambach2023b,kura2012a}. This, however, requires control of particle alignment coupled with a low-anisotropy material, specific particle size in the upper superparamagnetic range (typically less than 20 nm depending on material) \cite{zambach2023b}, a narrow size distribution, and ideally spherical particle shape \cite{zambach2023b,rowe2015a}.

Experimental and theoretical results show that high particle susceptibilities of $\chi_{\textrm{p}}>120$ (iron particles) and $\chi_{\textrm{p}}\sim30$ (iron-oxide particles) are possible for spherical superparamagnetic particles \cite{kura2012a,Zambach2025-DemagPaper,Zambach2025-MaterialPaper}. 
However, to date, it has not been possible to achieve nanocomposites with high susceptibility \cite{FeNi3_Article_Lu,yun2014a,yun2016a,yatsugi2019a,liu2005a,kura2012a,kura2014a,yang2018a,kin2016a,garnero2019a,hasegawa2009a,rowe2015a,Zambach2025-MaterialPaper}. Nanocomposites using well dispersed iron-oxide or Zn-ferrite particles report susceptibilities lower than 20 \cite{Zambach2025-MaterialPaper,yun2014a,yun2016a,rowe2015a,Luo_FeOx2024}. Nanocomposites that use high saturation materials (FeNi$_4$ or FeCo) achieve susceptibilities upwards of 60, but suffer from large power losses due to their high coercivity induced by aggregation \cite{yatsugi2019a,yang2018a}.

To access the full potential of nanocomposites as soft magnetic materials for inductor core materials, one must precisely control particle composition, size, shape, orientation, and dispersion \cite{zambach2023b,Zambach2025-DemagPaper,rowe2015a}. Aggregation of particles promotes the transition from the superparamagnetic state to the blocked state, resulting in large coercivity and low susceptibility \cite{zambach2023b,Durhuus2025a}. Thus, magnetic losses are increased by particle aggregation, too large particle sizes, and/or too wide particle size distribution \cite{Bima_Preprint,zambach2023b,Zambach2025-MaterialPaper}.

The susceptibility of uniaxial anisotropy superparamagnetic particles depends on the applied field direction relative to the particle's anisotropy axis \cite{zambach2023b,svedlindh1997a,raikher1974a,shliomis1993a}. The random orientation particle susceptibility $\langle\chi_{\textrm{p}}\rangle$ is calculated as the directional average \cite{zambach2023b}
\begin{equation}
    \langle\chi_{\textrm{p}}\rangle = \frac{1}{3}\chi_{\textrm{p}\parallel} + \frac{2}{3}\chi_{\textrm{p}\perp},
    \label{eq:ChiLin}
\end{equation}
with $\chi_{\textrm{p}\parallel}$ / $\chi_{\textrm{p}\perp}$ being the particle susceptibility where the anisotropy axis and applied field are parallel / perpendicular.
As $\chi_{\textrm{p}\parallel} >\langle\chi_{\textrm{p}}\rangle$ for most regimes \cite{zambach2023b}, one expects a two- to three-fold increase in susceptibility for parallelly aligned iron-oxide particles compared to the randomly orientated case \cite{zambach2023b,li2024a,ludwig2017a}. The relaxation process for the perpendicular direction is characterised by a shorter timescale compared to the parallel direction, giving rise to differences in the frequency dependence of $\chi_{\textrm{p}\parallel}$ and $\chi_{\textrm{p}\perp}$ \cite{zambach2023b,svedlindh1997a,raikher1974a,shliomis1993a}. Further details on the frequency dependence of the parallel and perpendicular particle susceptibilities can be found in the \hyperref[sect:theory]{Theory} section.

Experimental work on particle alignment in nanocomposites is limited with few studies on iron-oxide and Co particles \cite{ludwig2017a,li2024a,kura2014a}. In these studies, no morphological analyses were presented to access particle aggregation or interaction effects. Theoretical studies predict reduced power losses for nanocomposites containing aligned particles compared to the randomly orientated particles, but this remains to be experimentally verified \cite{zambach2023b}.

Interparticle interactions are predicted to increase nanocomposite susceptibility under certain conditions \cite{elfimova2019a,Goldfarb2026a}. For weakly interacting particles, nanocomposite susceptibility $\chi_{\textrm{nc}}$ increases super-linearly with particle susceptibility \cite{elfimova2019a}
\begin{equation}
    \chi_{\textrm{nc}} = f \chi_{\textrm{p}} \left( 1 +  \frac{1}{3}f\chi_{\textrm{p}}\right).
    \label{eq:ChiInt}
\end{equation}
Here $f$ is the volume fraction of particles in the nanocomposite and $\chi_{\textrm{p}}$ is the particle susceptibility. For the non-interacting case the nanocomposite susceptibility follows the particle susceptibility linearly \cite{carrey2011a,zambach2023b}.
Combining the increase in susceptibility by particle alignment and particle interactions of equation \eqref{eq:ChiInt} shows that nanocomposite susceptibility values as high as 90 should theoretically be possible, even when using iron-oxide particles \cite{zambach2023b,Zambach2025-MaterialPaper}. To our knowledge, no experimental investigation of particle interaction on susceptibility for aligned particles has been presented in the literature.

In this paper, we investigate susceptibility and power losses in nanocomposites containing well-dispersed field-aligned superparamagnetic 11$\pm$3 nm maghemite ($\gamma$-Fe$_2$O$_3$) nanoparticles. 
The dependences on alignment field strength and particle concentration were investigated, and small-angle scattering was used to quantify particle aggregation in the obtained nanocomposites.  Hysteresis and susceptibility measurements in both DC and AC conditions were used to link morphology, particle alignment, and interactions with susceptibility and losses in the high kHz range.
The degree of particle alignment was obtained by combined fitting of the susceptibility data from samples made with and without alignment field.

\section{Results}
\subsection*{Nanocomposite synthesis and characterisation}
\begin{figure}[bp!]
    \centering
    \includegraphics[width=1\columnwidth]{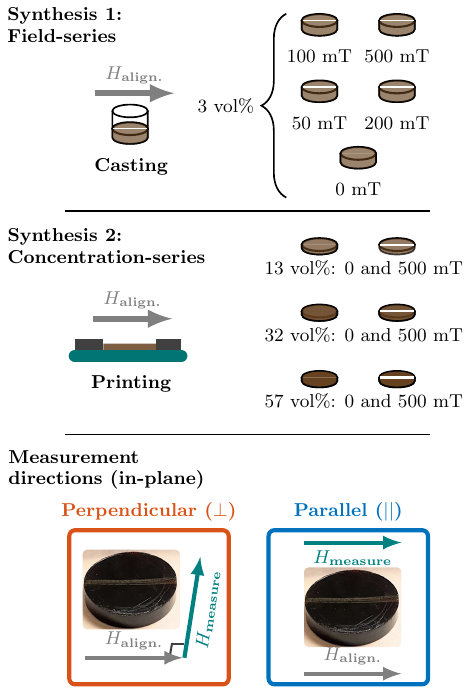}
    \caption{Illustration of nanocomposite synthesis and in-plane measurement directions on nanocomposite disk samples. Nanoparticle solution was mixed with PVA and cast into disk shape (field-series) or printed in layers, yielding flat disks (concentration-series). White markings indicate alignment field direction on sample.}
    \label{fig:OverviewComic}
\end{figure}
A stable aqueous solution of log-normal distributed $11\pm3$ nm diameter maghemite ($\gamma$-Fe$_2$O$_3$) nanoparticles was obtained by a polyol process as described in the experimental section and previous work \cite{Zambach2025-MaterialPaper}. The particles were well-dispersed in the solution due to electrostatic repulsion provided by a low pH. The concentrated nanoparticle solution (44.1 mg iron-oxide per ml) was mixed directly with poly-vinyl alcohol polymer (PVA) with volumes / masses given in the experimental section.

Two sample series were prepared, named field-series and concentration-series, see figure \ref{fig:OverviewComic}. For the field-series, containing samples with 3 $\pm$0.3 vol\% particles, the strength of the alignment field was 0, 50, 100, 200, and 500 mT.
In the concentration-series, the volume fraction was varied (13 vol\%, 32 vol\%, and 57 vol\%), and sample pairs were synthesized with either 0 mT or 500 mT alignment field. For both series, the alignment field was static, homogeneous, and applied by drying the samples in the pole gap of an electromagnet. The alignment field direction was marked on the sample disks, and magnetic characterisation was performed with measurement field parallel ($\parallel$) and perpendicular ($\perp$) to the alignment field direction in the plane of the disk samples, see figure \ref{fig:OverviewComic}.

The field-series (3 vol\%) samples were cast in one step, with several days drying time. The samples were placed in the alignment field after pipetting the particle-polymer solution into the cast. The concentration-series (13, 32, and 57 vol\%) were prepared by a layer-by-layer coating process, resembling micro-fabricational methods. Each layer was deposited outside of the alignment field, the sample was then placed inside the alignment field while the the film was drying (20 min per layer).
\begin{figure}[tp!]
    \centering
        \includegraphics[width=1\columnwidth]{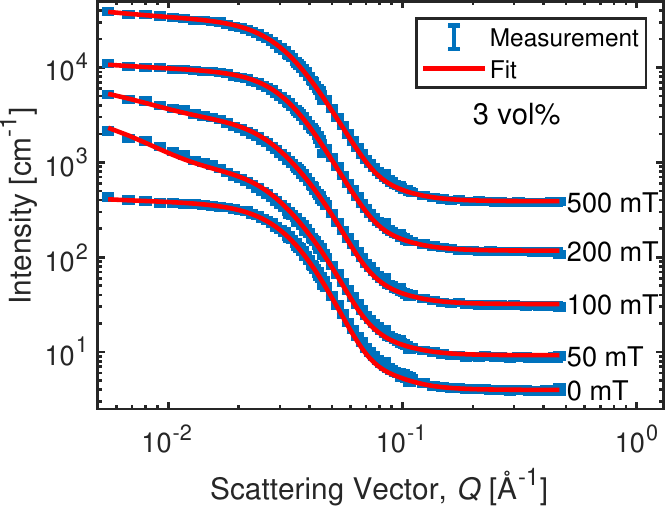}
        \caption{Small-angle neutron scattering (SANS) data for field-series samples containing 3 vol\% 11$\pm$3 nm maghemite particles together with two-model-fit. Scattering intensity was scaled with $10^{x}$, $x\in[0,1,..]$ for better visibility.}
        \label{fig:SANS}
\end{figure}

Both preparation methods yielded solid, stable polymer-like disks. The field-series resulted in samples with diameters of 12 mm and height of around 1.8 mm (see example in figure \ref{fig:OverviewComic}), and the concentration-series yielded samples of 8.4 mm diameter and a height between $0.3-1$ mm. 

Small angle scattering was performed on all samples to investigate particle aggregation. The field-series was probed using small angle neutron scattering (SANS) \cite{alm2021a}, and the concentration-series was probed by small angle X-ray scattering (SAXS) \cite{MaxIV_CoSaxs}. Small-angle data was obtained for a $Q$-range of 0.005-0.5 Å$^{-1}$ (SANS) and 0.003-0.3 Å$^{-1}$ (SAXS). The scattering data was fitted by a two-model-fit, modelling single particles (spherical form-factor multiplied by hard-sphere structure-factor) and aggregates (spherical form-factor multiplied with fractal structure factor), as described in the supplementary material and previous work \cite{Zambach2025-MaterialPaper}.
\begin{figure}[tp!]
    \centering
    \includegraphics[width=1\columnwidth]{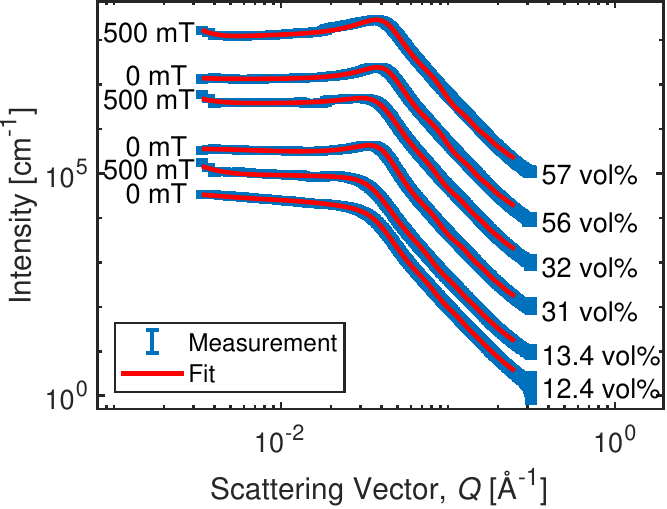}
    \caption{Small-angle x-ray scattering (SAXS) data for concentration-series samples with (500 mT) and without alignment field (0 mT) together with two-model-fit. Scattering intensity was scaled with $10^{x}$, $x\in[0,1,..]$ for better visibility.}
    \label{fig:SAXS}
\end{figure}

Figure \ref{fig:SANS} shows SANS data for the field-series, together with the two-model-fit. The scattering profile resembles the characteristic spherical nanoparticle scattering with $Q^{-4}$ dependence in medium $Q$-range, and relatively flat intensity profile in the low $Q$-range for the 0, 200, and 500 mT alignment fields. An increase in intensity in the low $Q$ signal is seen for 50 and 100 mT alignment fields.
The fits to the SANS data are seen to reproduce the scattering well with particle diameters of around 9.0$\pm$2.8 to 10.2$\pm$2.6 nm. Slight deviation from measured intensity is seen for the $Q$-range of 0.07-0.1, possibly due to non-sufficient polymer background subtraction. Full list of fitting parameters can be seen in the supplementary material. 

Analysis of the SANS fitting shows that for the 0 mT, 200 mT, and 500 mT alignment field samples 94-98 vol\% of the particles are in the single particle configuration, with the remaining 2-6 vol\% being in round aggregates of 3-6 particles. The 50 mT and 100 mT alignment field samples show lower amount of single particle configuration (47-61 vol\%), but with the remaining 39-53 vol\% being in aggregates of only 2-3 particles, see figure S1 in the supplementary material.

Analysis of the 2D scattering data shows that for the field-series the particles seemed elliptical with 1 nm longer particle diameter along the alignment field direction. For the 50 mT and 100 mT alignment fields, aggregates did not orient along the alignment field, suggesting that $\leq$100 mT is insufficient to keep the particles oriented against Brownian agitation.
A lower alignment field also resulted in seemingly more aggregation with smaller aggregates. This alignment field dependence should be investigated further.

Figure \ref{fig:SAXS} shows the SAXS data for the concentration-series, together with two-model-fits. The scattering profile for the concentration-series (layered coatings) is much more flat in the low $Q$-range compared to the cast samples. A minimal increase of intensity is seen for the lowest $Q$-range. The concentration-series have higher volume fraction of particles than the field-series (3 vol\%), and thus the characteristic hard-sphere interaction peak around $Q\approx0.04$ Å$^{-1}$ is more pronounced. The particle diameters from two-model-fits in the concentration-series vary within uncertainty from 9.4$\pm$3.6 for low concentrations to 7.2$\pm$2.2 at high concentration.

Further analysis of the SAXS fitting parameters shows that the concentration-series overall has a higher fraction of particles in the single particle configuration ($>$98 vol\% single particles) for samples with and without alignment field compared to the field-series samples. This is also apparent from the flatter low $Q$ profile of the scattering intensity seen in figure \ref{fig:SAXS} compared to the scattering profiles of the field-series of figure \ref{fig:SANS}. The lower amount of aggregation is in agreement with the 0 mT and 500 mT results on the 3 vol\% sample from the field series and potentially further supported by the faster drying of the concentration-series. For details on the fitting parameters and their analysis, see the supplementary material.

To date, there is limited literature on nanoparticle alignment in magnetic dense nanocomposites. Often materials use larger particles \cite{kura2014a}, or when superparamagnetic particles are used, the systems studied use low concentration of particles \cite{svedlindh1997a}. In the literature, no quantification of aggregation was found for magnetic nanocomposites. Transmission electron microscopy (TEM) is generally used to show the aggregation state, as seen in \cite{kura2014a}, where FeCo nanoparticles were cured in a polystyrene matrix, but the method does not give quantitative estimates of particle aggregation. The low degree of particle aggregation for the composites presented in this paper is probably a consequence of the difference in synthesis from previous nanocomposites presented in the literature. We have utilised surface charging, faster drying times, and smaller particle moment (and therefore less magnetic attraction) compared to other field-aligned nanoparticle materials \cite{kura2014a}.

\subsection*{Magnetic characterisation}
Magnetic characterisation in the form of vibrating sample magnetometry (VSM), AC-susceptibility, and High-frequency hysteresis loop measurements was performed across all samples. These measurements were performed along the diameter in the plane of disk-shaped samples, parallel ($\parallel$) and perpendicular to the alignment field axis for samples synthesised in alignment field, see figure \ref{fig:OverviewComic}. For samples where no alignment field was used (0 mT), no directional dependence of the susceptibility on the measuring direction along the disk diameter was found.
\begin{figure}[bp!]
        \centering
        \includegraphics[width=1\columnwidth]{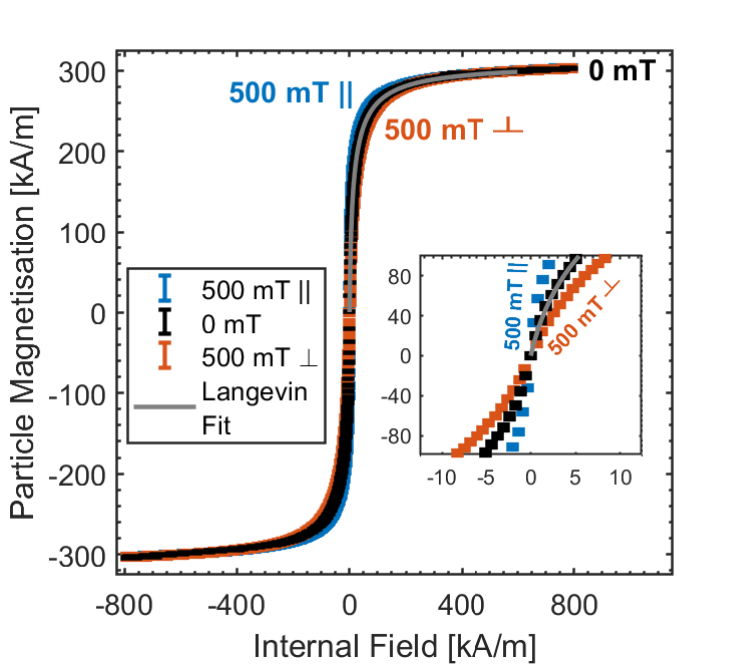}
        \caption{Normalised VSM hysteresis curves for the 56 vol\% sample with and without alignment field together with Langevin fit. Measurements were obtained at room temperature. For the samples synthesised with 500 mT alignment field, hysteresis for the direction measuring parallel ($||$) and perpendicular ($\perp$) to the alignment field direction are shown. VSM results and fitting parameters are listed in table \ref{tab:VSMResults}}
        \label{fig:VSM_HMA}
\end{figure}

Figure \ref{fig:VSM_HMA} shows the hysteresis curves, normalized to nanoparticle volume, obtained by VSM for the samples containing 57 vol\% particles synthesised with and without alignment field together with a theoretical prediction for randomly oriented superparamagnetic particles (Langevin fit). Hysteresis curves for all other samples can be seen in the supplementary material. The Langevin model is presented in the \hyperref[sect:theory]{Theory} section.
All samples show no coercivity (i.e. it is below detection limit of 8 A/m), and Langevin fitting diameters between 7.8$\pm$4.0 nm and 9.6$\pm$4.0 nm. 
The results are summarised in table \ref{tab:VSMResults}. Langevin fitting volume fraction agree with volume fraction found with measured particle saturation magnetisation (303 kA/m). We note that the Langevin model fits well even for the most concentrated sample (57 vol\%) with particle distances below 1 nm. The observed decrease in fitting diameter and increased particle size distribution width for concentrated samples are small but likely due to particle interactions \cite{elfimova2019a,Goldfarb2026a}.

The fitting diameter from the Langevin model was seen to be slightly smaller than expected from TEM (11$\pm$3), but but in agreement with SANS/SAXS fitting results. Differences from TEM, VSM and small-angle scattering can be explained by differences in number- vs volume-weighting of the log-normal distribution. 
\begin{table}[tp!]
\centering
 \caption{Results from VSM hysteresis measurements at room temperature. Nanoparticle content was obtained by measured saturation magnetisation (303 kA/m). Fitting diameter was obtained from fitting of Langevin with log-normal distributed particle diameters for samples with no alignment field.}
 \label{tab:VSMResults}
 \begin{ruledtabular}
    \begin{tabular}{ccc|ccc}
    NP content & $H_{\textrm{align.}}$ & Orien- & $H_{\textrm{c}}$ & Fit dia. & Sample \\\relax
    [Vol\%] & [mT] & tation & [A/m] & [nm] & susc. $\chi_{\textrm{nc}}$\\
    \hline
    $3.0\pm0.2$ & 0 &  & $3\pm8$ & $9.4\pm3.1$ & $0.40\pm0.04$ \\
    $2.7\pm0.2$ & 50 & $||$ & $4\pm8$ & - & $0.63\pm0.07$ \\
    $2.7\pm0.2$ & 50 & $\perp$ & $7\pm8$ & - & $0.30\pm0.05$ \\
    $2.8\pm0.2$ & 100 & $||$ & $4\pm8$ & - & $0.64\pm0.06$ \\
    $2.8\pm0.2$ & 100 & $\perp$ & $4\pm8$ & - & $0.30\pm0.04$ \\
    $2.8\pm0.2$ & 200 & $||$ & $4\pm8$ & - & $0.54\pm0.05$ \\
    $2.8\pm0.2$ & 200 & $\perp$ & $4\pm8$ & - & $0.33\pm0.04$ \\
    $2.7\pm0.2$ & 500 & $||$ &  $6\pm8$ & - & $0.55\pm0.06$ \\
    $2.7\pm0.2$ & 500 & $\perp$ & $7\pm8$ & - & $0.29\pm0.04$ \\
    \hline
    $12.4\pm0.2$ & 0 &  & $6\pm8$ & $9.6\pm4.0$ & $4.2\pm0.2$ \\
    $13.4\pm0.2$ & 500 & $||$ &  $9\pm8$ & - & $10.6\pm0.5$ \\
    $13.4\pm0.2$ & 500 & $\perp$ & $0\pm8$ & - & $2.1\pm0.1$ \\
    $30.5\pm1.0$ & 0 &  & $3\pm8$ & $9.0\pm4.1$ & $11.5\pm0.9$ \\
    $31.8\pm1.0$ & 500 & $||$ &  $5\pm8$ & - & $28.0\pm3.0$ \\
    $32.3\pm1.0$ & 500 & $\perp$ & $4\pm8$ & - & $7.5\pm0.7$ \\
    $56.1\pm1.0$ & 0 &  & $-1\pm8$ & $7.8\pm4.0$ & $21.2\pm2.1$ \\
    $56.5\pm1.0$ & 500 & $||$ &  $8\pm8$ & - & $50.1\pm5.1$ \\
    $57.0\pm1.0$ & 500 & $\perp$ & $4\pm8$ & - & $11.5\pm1.0$
  \end{tabular}
 \end{ruledtabular}
\end{table}

The measured susceptibility for 0 mT alignment field samples are between the parallel and perpendicular cases for samples where alignment field was used. The susceptibility of the 3 vol\% sample without field alignment is equal to the weighted sum of the parallel and the perpendicular case, as expected for the case of negligible interactions, see eq. \eqref{eq:ChiLin} but this is not the case for the more concentrated samples. The reported susceptibility of the 57 vol\% sample in the parallel case ($\chi_{\textrm{nc}} = 50\pm5$) is among the highest susceptibilities reported for nanocomposite systems, including systems of metallic particles \cite{FeNi3_Article_Lu,yun2014a,yun2016a,yatsugi2019a,liu2005a,kura2012a,kura2014a,yang2018a,kin2016a,garnero2019a,hasegawa2009a,rowe2015a,Zambach2025-MaterialPaper}.
\begin{figure}[tp!]
        \centering
        \includegraphics[width=1\columnwidth]{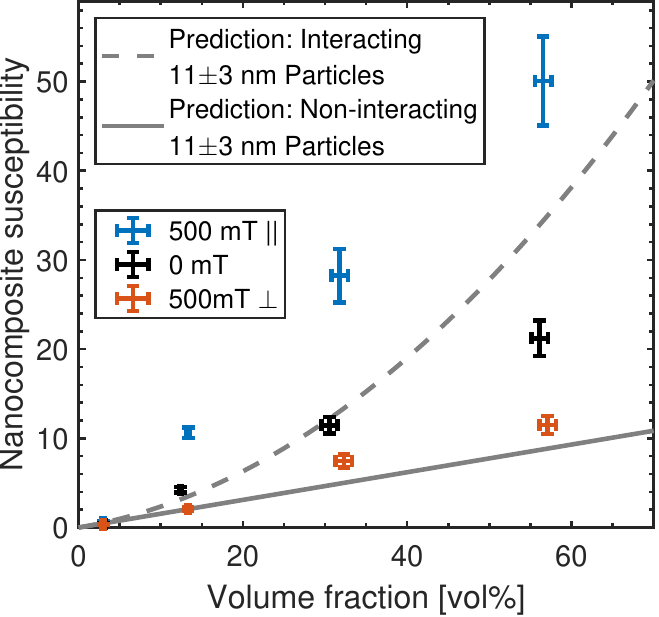}
        \caption{Nanocomposite susceptibility found from VSM as function of volume fraction for samples synthesised with 500 mT alignment field and without alignment field. Predicted susceptibility for randomly oriented, non-interacting linear model (eq. \eqref{eq:ChiLin}) and for randomly oriented, weakly interacting particles (eqs. \eqref{eq:ChiLin} and \eqref{eq:ChiInt}) for 11$\pm$3 nm maghemite particles are shown as full / dotted lines.}
        \label{fig:VSM_Susc}
\end{figure}
\begin{figure*}[tp!]
        \centering
        \begin{minipage}[t]{1\columnwidth}
            \centering
            \includegraphics[width=1\columnwidth]{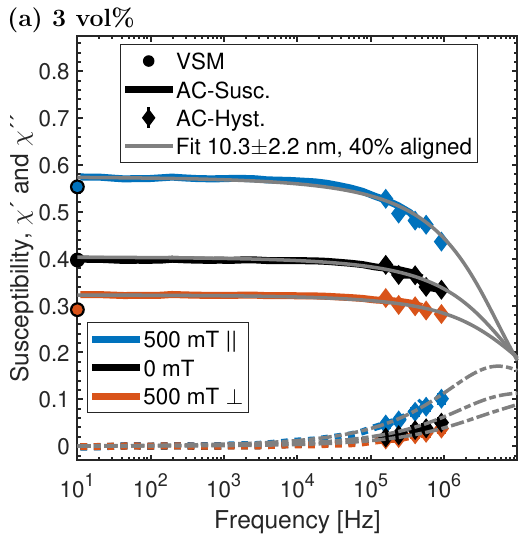}
        \end{minipage}
        \hfill
        \begin{minipage}[t]{1\columnwidth}
            \centering
            \includegraphics[width=1\columnwidth]{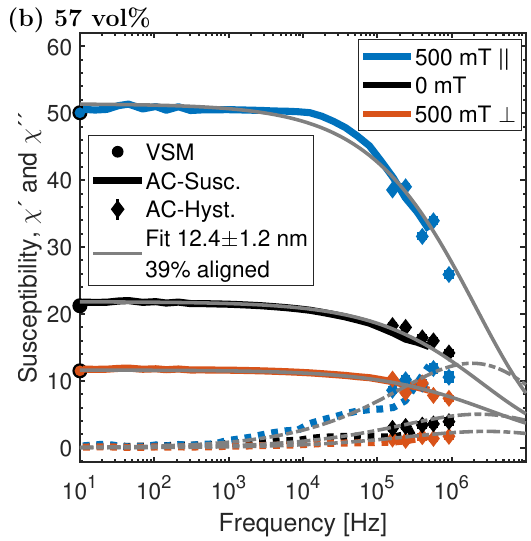}
        \end{minipage}
        \caption{AC-susceptibility for nanocomposite sample pairs containing 3 vol\% (a) and 57 vol\% particles (b), together with Debye model fit including interactions. In- and out-of-phase components $\chi'$ and $\chi''$ shown as solid/dashed lines. Data for samples with and without alignment field are fitted simultaneously for each sample pair, for fitting results see table \ref{tab:ACResults}. AC-susceptibility data and fits of the other samples can be found in the supplementary material, for theoretical fitting expressions see the \hyperref[sect:theory]{Theory} section.}
        \label{fig:ACSusc_Field}
\end{figure*}

Figure \ref{fig:VSM_Susc} shows susceptibilities from VSM data for all nanocomposites, together with interacting \eqref{eq:ChiInt} and non-interacting Langevin susceptibility predictions for randomly oriented 11$\pm$3 nm maghemite particles. We find that the susceptibility of the nanocomposites increases super-linearly for particle fractions above 3 vol\%. This is likely due to particle interactions. The mean-field prediction of equation \ref{eq:ChiInt} describes the susceptibility well, except for the 57 vol\% samples, which is lower than expected. The interaction model is based on relatively weak interactions and might not be a suitable fit for the denser 57 vol\% case \cite{elfimova2019a}. 
The susceptibility of the samples synthesised with alignment field are seen to be considerably higher/lower than that for the same concentrations synthesised without alignment field. The increase in susceptibility for the parallel case is not consistent with the principles eq. \eqref{eq:ChiLin} alone. Only the combination of particle alignment and interaction from eqs. \eqref{eq:ChiLin} and \eqref{eq:ChiInt} explains the measured susceptibilities.

The susceptibility of the field-series is rather stable wrt. alignment field. Only slight relative increase/decrease in susceptibility is seen for the field aligned cases measured parallel/perpendicularly, indicating that lower alignment field results in bigger difference between the two directions.

The nanocomposite samples were analysed by AC-susceptometry in the frequency range from 11 Hz to 922 kHz, using the same three different applied field configurations as for the VSM measurements, see figure \ref{fig:OverviewComic}. Measurements were performed on a dedicated AC-susceptometer (10 Hz to 500 kHz) and on an AC-hysteresis setup (166 to 922 kHz) \cite{TVeile2025}, see experimental section. Figure \ref{fig:ACSusc_Field} shows the resulting AC-susceptibility measurements for the 3 vol\% samples and 57 vol\% samples for 0 mT and 500 mT alignment field. AC-susceptibility data for the other samples can be found in the supplementary material. 
We find that the susceptibility is constant for frequencies up to around 20 kHz, 40 kHz, and 100 kHz for the parallel, 0 mT, and perpendicular cases. Above these frequencies the in-phase component of the susceptibility starts to decline, accompanied by an increase in out-of phase component. The difference in frequency at which the susceptibility starts to decrease can be explained by the difference in relaxation time parallel and perpendicular to the particle easy axis in superparamagnetic particles \cite{zambach2023b,svedlindh1997a}, see the \hyperref[sect:theory]{Theory} section.

The particle size, saturation magnetisation, uniaxial anisotropy constant, superparamagnetic attempt time, and degree of particle alignment can be extracted by fitting the AC-susceptibility for the 0 mT, parallel and perpendicular cases simultaneously \cite{zambach2023b}. The degree of alignment is modelled by fractions of particles perfectly aligned, with the remainder of the particles being in the random direction configuration. The fitting model is thus a combination of the directional Debye-model \cite{zambach2023b,svedlindh1997a,raikher1974a,shliomis1993a} and interaction effects of eq. \eqref{eq:ChiInt}, see also the \hyperref[sect:theory]{Theory} section. Table \ref{tab:ACResults} shows the fit results from the AC-susceptibility data in figure \ref{fig:ACSusc_Field} and all other samples.

For low volume fractions the fitting diameter fits well with the found diameter from VSM data. The effective diameter increases slightly with particle volume fraction due to interaction effects as discussed earlier. The effective particle anisotropy constants ($K_{\textrm{u}}$) found by fitting are around 10-35 kJ/m$^3$, typical values for maghemite \cite{rivas2022a}. 
For the field-series we find that increasing the alignment field decreases the alignment fraction.

\begin{figure}[bp!]
        \centering
        \includegraphics[width=1\columnwidth]{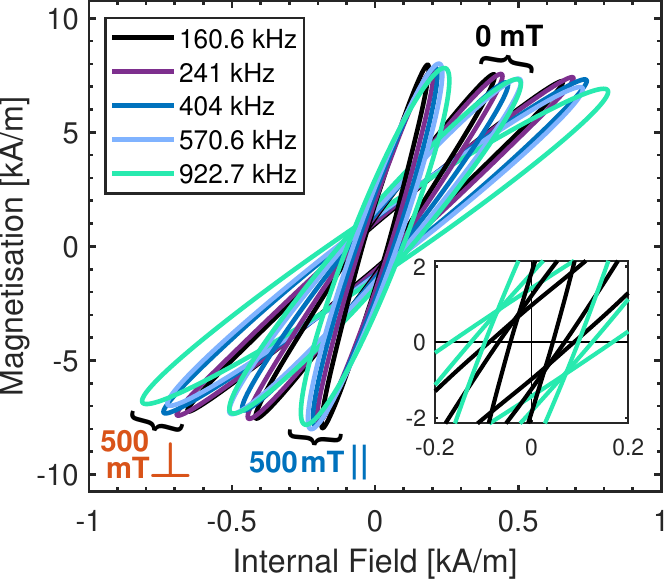}
        \caption{High frequency hysteresis loops for the 57 vol\% sample pair at a fixed 10 mT internal $B$-field amplitude. Data shown for samples dried in a 0 mT and 500 mT alignment field, measured parallel and perpendicular to the alignment field direction. Inset: Zoom in for the lowest and highest frequencies.}
        \label{fig:ACHyst_Field}
\end{figure}

High frequency hysteresis measurements in the frequency interval of 160.6 to 922.7 kHz were performed while scaling the applied field such that the internal $B$-field is 10 mT. This simulates the use case of an inductor core material, where losses and performance are evaluated per generated flux in B-H loops. Figure \ref{fig:ACHyst_Field} shows the resulting high frequency hysteresis loops for the 57 vol\% samples in the parallel and perpendicular configuration, together with the results for no alignment field. Hysteresis loops for the other samples in the concentration-series can be found in the supplementary material.
\begin{table*}
    \centering
    \caption{Fitting results for simultaneously fitting of AC-susceptibility of sample pairs synthesised with and without alignment field, for the three direction configurations, as seen in figure \ref{fig:ACSusc_Field}. Particle saturation magnetisation ($M_{\textrm{s}}$), effective uniaxial particle anisotropy ($K_{\textrm{u}}$), superparamagnetic attempt time ($\tau_0$), and fraction of particles that are aligned to the alignment field direction ($\varphi$). For detailed description of the fitting model see \hyperref[sect:theory]{Theory} section.}
    \label{tab:ACResults}
    \begin{ruledtabular}
        \begin{tabular}{cc|cccccc}
        NP content & $H_{\textrm{align.}}$ & Fit NP Content &  Fit diameter & Fit $M_{\textrm{s}}$ & Fit $K_{\textrm{u}}$ & Fit $\tau_0$ & Fit alignment \\\relax
        [Vol\%] & [mT] & [Vol\%] & [nm] & [kA/m] & [kJ/m$^3$] & [ns] & fraction $\varphi$ [\%]\\
        \hline
        $3$ & 0, 50 & 3.4 &  $9.7\pm2.3$ & 333 & 8.4 & 13.8 &  63 \\
        $3$ & 0, 100 & 3.4 &  $10.1\pm2.2$ & 331 & 9.7 & 7.7 & 61 \\
        $3$ & 0, 200 & 3.4 &  $10.0\pm2.3$ & 333 & 8.4 & 12.2 &  32 \\
        $3$ & 0, 500 & 3.4 &   $10.3\pm2.2$ & 329  & 10.5 & 7.9 & 40 \\
        \hline
        $13$ & 0, 500 & 15.2 &   $11.9\pm2.5$ & 329 & 12.4 & 6.2 & 62\\
        $32$ & 0, 500 & 37.0 &   $12.3\pm1.7$ & 326 & 21.4 & 0.9 & 42\\
        $57$ & 0, 500 & 61.0 &   $12.4\pm1.2$ & 314 & 34.5 & 0.07 & 39
        \end{tabular}
    \end{ruledtabular}
\end{table*}

The high frequency hysteresis loops in figure \ref{fig:ACHyst_Field} show that susceptibility decreases with frequency while coercivity increases as expected respectively from the in-phase and out-of-phase components of the susceptibility in figure \ref{fig:ACSusc_Field}. Hence, to reach the same 10 mT internal $B$-field, a larger applied field is needed with increasing frequency. Due to the larger susceptibility in the parallel and random cases, lower external applied fields are needed compared to the perpendicular case. 
In principle, the random configuration is expected to have the highest out-of-phase component and thus highest coercivity \cite{zambach2023b,svedlindh1997a}, however, this only holds for 100\% alignment. For the case shown in figure \ref{fig:ACHyst_Field}, where alignment is roughly 39\%, we see that the 61\% randomly oriented particles increase the coercivity of the perpendicular case. For samples with higher alignment, the coercivity for the perpendicular direction is smaller than for the parallel and random directions, as expected \cite{zambach2023b,svedlindh1997a} (see supplementary material). The decrease in susceptibility for increasing frequency is stronger for the perpendicular and random cases, as expected from the AC-susceptibility.
\begin{figure}[bp!]
        \centering
        \includegraphics[width=1\columnwidth]{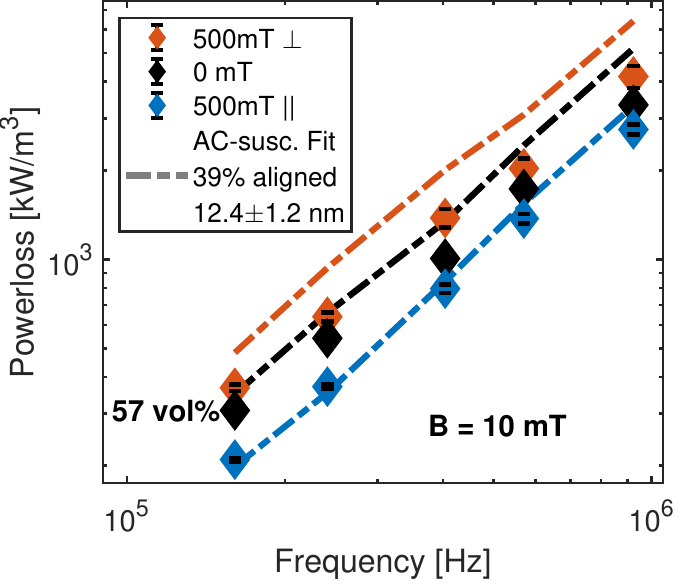}
        \caption{Power loss at a 10 mT internal $B$-field amplitude for the 57 vol\% sample pair based on measurements seen in figure \ref{fig:ACHyst_Field}. Predicted losses are based on fitting results of Debye model with interactions to AC-susceptibility data, see figure \ref{fig:ACSusc_Field} and table \ref{tab:ACResults}. For detail on the power loss calculation see \hyperref[sect:theory]{Theory} section.}
        \label{fig:Losses_Field}
\end{figure}

Figure \ref{fig:Losses_Field} shows the power losses per sample volume for the 57 vol\% sample, calculated based on the measured high frequency hysteresis loops of figure \ref{fig:ACHyst_Field}. The theoretical power loss predictions in figure \ref{fig:Losses_Field} are based on the fits of the complex susceptibilities in figure \ref{fig:ACSusc_Field} (table \ref{tab:ACResults}), see also the \hyperref[sect:theory]{Theory} section. Plots of measured power losses for the other samples can be seen in the supplementary material. For the 57 vol\% sample, the power losses are smallest for the sample synthesised with alignment field when applying the AC-measurement field parallel to the alignment field direction, and highest for the direction perpendicular to the applied field. One would expect the 0 mT sample to have highest power loss per induced flux when compared to the case where all particles are aligned \cite{zambach2023b}, both in the parallel and perpendicular cases. Since alignment here is only around 39 \% the order is changed, as the decrease in out-of-phase component of the perpendicular case is less than with 100\% alignment. Therefore, here the perpendicular case has the highest losses, which agrees with theoretical prediction also shown in figure \ref{fig:Losses_Field}. For samples with higher alignment the perpendicular direction has lower losses than the random and parallel case, see for example the 13 vol\% sample with 500 mT alignment field in the supplementary material.

Figure \ref{fig:Losses_Field} shows that the power losses increase with frequency to the power of ca. 1.2. The theoretical predictions increase with the frequency to the power of 1.3, 1.6, and 1.5 respectively. The lower than expected power losses at high frequencies can be explained by heating up of the sample during the high frequency hysteresis measurements. The increase in sample temperature results in a decrease in superparamagnetic relaxation time, i.e. a decrease in out-of-phase component of the susceptibility \cite{zambach2023b}. Theoretical predictions agree with measurements even for high volume fractions, as the effect of interactions on the relaxation time seem to be absorbed in anisotropy and attempt time fitting constants (cf. table \ref{tab:ACResults}). The decrease in effective particle size found from small angle scattering and Langevin fitting is not found in the AC fitting, which could indicate that interaction effects are also absorbed in the particle diameter fitting parameters for the AC-susceptibility.

Power losses of the here presented nanocomposites are high compared to ferrite materials used today for power converters. The state-of-the-art PC200 ferrite from TDK for example shows power losses of around 50-200 kW/m$^3$ at 30 mT and 1 MHz \cite{sanusi2022a,TDK}. Power losses of such ferrites however, increase more strongly with frequency, with exponents of roughly 2 \cite{sanusi2022a,TDK}. Thus, for low MHz frequencies the presented nanocomposite cannot compete with the state-of-the-art ferrite materials. For frequencies above some MHz however, nanocomposites have an advantage over ferrites. Optimisation of the magnetic properties of nanocomposites will require further tuning of magnetic particle characteristics, for example refined particles size distribution \cite{zambach2023b}. Also, magnetic soft materials with high saturation, like FeNi$_x$, FeCo, and Fe have shown to be potential candidates with lower losses per induced $B$-field \cite{zambach2023b,kura2012a}.

\section{Conclusion}
This study proves experimentally that alignment of superparamagnetic particles by an alignment field can increase magnetic susceptibility and decrease power losses in nanocomposite materials. Nanocomposite susceptibility increases super-linearly with particle fraction, and shows two- to three-fold rise from synergy between particle alignment and particle interactions.

Specifically, nanocomposites containing well-dispersed superparamagnetic 11$\pm$3 nm maghemite ($\gamma$-Fe$_2$O$_3$) particles with particle alignment of up to 62\% were prepared. A low degree of aggregation was achieved with more than 98 vol\% of particles in a single particle configuration by using electrostatically charged nanoparticles, large homogeneous alignment fields (200 mT and 500 mT) and a fast drying layer-by-layer printing deposition approach. 

The nanocomposites volume susceptibility followed predictions for interacting particles also for field aligned particles, experimentally confirming that theoretical prediction for weakly interacting particles also holds for aligned particles.

AC-susceptibility and high frequency hysteresis measurements demonstrate that the dependence of susceptibility on the direction of the particle easy axis towards the applied field direction can be used in material design as predicted in \cite{zambach2023b}. Substantial increase in volume susceptibility from 21 to 50 is possible by drying the 57 vol\% $\gamma$-Fe$_2$O$_3$ nanocomposite in a homogenous 500 mT alignment field while well-dispersiveness is maintained. A nanocomposite susceptibility of 50 is noticeable and makes the material relevant for power electronics \cite{zambach2023b}. The increase in susceptibility also has direct effect on power losses, which, at a fixed internal $B$-field of 10 mT, decrease by 25-50\%. Partial alignment of particles resulted in increased losses for the case where the applied/measurement field is perpendicular to the alignment field compared to the case without alignment field. AC-susceptibility measurements up to 922.7 kHz further show that superparamagnetic relaxation does not change dramatically even for high volume fractions of particles, possibly due to the low magnetic moment of the used particles.

The material presented here has a higher susceptibility than other iron-oxide containing nanocomposites, similar to materials containing FeCo and FeNi particles \cite{kura2014a,FeNi3_Article_Lu} and high frequency ferrites \cite{Fairite}. Although the presented nanocomposite shows higher losses than state-of-the-art ferrites in the kHz range, the power losses depend on frequency with a lower exponent than for ferrites (1.2 vs 2) \cite{sanusi2022a,TDK,Fairite}. This, together with predictions for an optimised particle size distribution to mitigate higher order losses \cite{Zambach2025-MaterialPaper}, shows that nanocomposites are promising candidates for MHz inductor core materials.

\section{Theory}
\label{sect:theory}
\subsection*{\textbf{Langevin Function}}
Magnetization data for samples synthesized in no alignment field was fitted with a Langevin function,
\begin{equation}
    M(H) = f M_{\textrm{s}} \left( \coth{\frac{\mu_0 M_{\textrm{s}} H V}{k_B T}} - \frac{k_B T}{\mu_0 M_{\textrm{s}} H V} \right).
    \label{eq:Langevin}
\end{equation}
Here $M$ is the volume magnetization of the nanocomposite, $H$ is the internal applied field, $f$ is the volume fraction of particles, $M_{\textrm{s}}$ is the saturation magnetization of the particle, and $T$ is the temperature. The Langevin model does not include interactions.

Polydispersity for all magnetic models was handled by volume weighting the fitting functions by the volume weighted log-normal distribution \cite{carrey2011a}.

\subsection*{Directional Debye-model with Interactions}
The particle susceptibility is calculated by a combination of the Debye-model including separate relaxation times for the parallel and perpendicular directions \cite{zambach2023b,svedlindh1997a,raikher1974a,shliomis1993a}. The parallel and perpendicular particle susceptibilities are
\begin{subequations}
\begin{align}
    & & \chi_{\textrm{p,}||} &= \epsilon_{\textrm{M}}\frac{R'}{R} \frac{1}{1+i\omega \tau_{||}}\\
    &\text{with} & \tau_\parallel &= \begin{cases}\tau_0\,2 \frac{R'}{R-R'}
    \\
    \tau_0 \frac{\sqrt{\pi}\exp(\epsilon_{\textrm{k}})}{2 \epsilon_{\textrm{k}}^{3/2}}
    \end{cases}
    \quad &\begin{array}{l}
        \textrm{for } \epsilon_{\textrm{k}}\leq2,\\[3pt]
        \textrm{for } \epsilon_{\textrm{k}} > 2,
    \end{array}
\end{align}
\label{eq:ChiPara}
\end{subequations}
and
\begin{subequations}
\begin{align}
    & & \chi_{\textrm{p,}\perp} &= \frac{\epsilon_{\textrm{M}}}{2}\left[1 - \frac{R'}{R}\right]\frac{1}{1+i\omega \tau_{\perp}} \\
    &\text{with} &
    \tau_\perp &= \tau_0\, 2 \frac{R - R' }{R + R'} 
    &\textrm{for all } \epsilon_{\textrm{k}}
\end{align}
\label{eq:ChiPerp}
\end{subequations}
In the above $\tau_0$ is the attempt time, $\epsilon_{\textrm{k}} = \frac{K_{\textrm{u}}V}{k_B T}$ is the weighted anisotropy barrier with the effective uniaxial anisotropy constant $K_{\textrm{u}}$, and $\epsilon_{\textrm{M}}=\frac{\mu_0 V M_{\textrm{s}}^2}{k_B T}$. The anisotropy weighting factors $R$ and $R'$ are
\begin{subequations}
    \begin{align}
        R &= \int_0^1 \exp\left( \epsilon_{\textrm{k}} x^2 \right) \textrm{d}x,\\
        R' &= \int_0^1 x^2 \exp\left( \epsilon_{\textrm{k}} x^2 \right) \textrm{d}x.
    \end{align}
\end{subequations}
The random orientation particle susceptibility is found by \eqref{eq:ChiLin}.

The susceptibility for a mixture of aligned and random oriented particles measured parallel and perpendicular to the alignment orientation is calculated as
\begin{subequations}
\begin{align}
    \chi_{\textrm{p,partial}||}(\varphi) &= \varphi \langle\chi_{\textrm{p}}\rangle + (1-\varphi)\chi_{\textrm{p,}||},
    \label{eq:ChiAlignPara}\\
    \chi_{\textrm{p,partial}\perp}(\varphi) &= \varphi \langle\chi_{\textrm{p}}\rangle + (1-\varphi)\chi_{\textrm{p,}\perp},
    \label{eq:ChiAlignPerp}
\end{align}
\label{eq:ChiMix}
\end{subequations}
where $\varphi$ is the fraction of aligned particles.

For interacting particles the susceptibility of the nanocomposite is then found by using eqs.\eqref{eq:ChiLin} or \eqref{eq:ChiMix} in eq. \eqref{eq:ChiInt} to include interactions. Sample-pair data for samples synthesised with and without alignment field were fitted simultaneously.

\subsection*{Power loss prediction}
Power losses in the small field regime for sinusoidal fields are predicted by 
\begin{equation}
    P(\nu) = \nu \mu_0 \oint \mathbf{M} d\mathbf{H} = \pi \nu \mu_0 H^2\chi''(\nu),
\end{equation}
where $\nu$ is the frequency, $H$ is the applied field amplitude, and $\chi''$ is the out-of-phase component of the complex susceptibility $\chi = \chi'-i\chi''$

\section{Experimental Section}

\paragraph*{\textbf{Nanoparticle synthesis and characterisation}}
Concentrated aqueous solution containing maghemite ($\gamma$-Fe$_2$O$_3$) nanoparticles with log-normal distributed diameter of 11$\pm$3 were synthesised by a polyol process as described in previous work \cite{Zambach2025-MaterialPaper}. The particle saturation magnetisation was measured to be 303 kA/m and particle sizes were measured by TEM.

\paragraph*{\textbf{Nanocomposite synthesis and characterisation}}
Nanocomposites were prepared by adding PVA polymer (9000-1000 M$_\textrm{w}$) directly to the particle solution by the amounts given in table \ref{tab:CompositeVol}. The resulting solution was stirred for 6 hours, until the polymer was dissolved completely in the solution.
\begin{table}[bp!]
\centering
 \caption{Volume of nanoparticle solution and mass of polyvinyl alcohol (PVA) polymer used for nanocomposite synthesis.}
 \label{tab:CompositeVol}
 \begin{ruledtabular}
  \begin{tabular}{cc}
    \textrm{NP Content} & \textrm{Polymer mass per particle}\\
    \textrm{[vol\%]} & \textrm{solution [mg/ml]}\\
    \colrule
    3  & 460 \\
    13 & 150 \\
    32 & 50 \\
    57 & 25
  \end{tabular}  
\end{ruledtabular}
\end{table}

Solution for the 3 vol\% case was filled into cylindrical casting moulds and dried at 40 degrees for 4-5 days. One sample was dried without an alignment field, whereas 4 samples were dried in a homogenous magnetic field from an electromagnet with the field parallel to the diameter of the mould cylinder, as seen in figure \ref{fig:OverviewComic}. Alignment field strength was measured by use of the Gaussmeter of a Lake Shore Cryotronics 7407-series VSM and was set to either 50, 100, 200, or 500 mT. Alignment field direction was marked on the samples (white line in figure \ref{fig:OverviewComic}). Finished samples were disk shaped with diameter of 12 mm and height of 1-2 mm.

Solutions for the 13, 32 and 57 vol\% samples were spread in a layer-by-layer manual printing into a round mould, as described in \cite{Zambach2025-MaterialPaper}. 20 microliter solution was deposited, then placed in the homogenous alignment field and dried for 20-30 minutes. The process was repeated, until disks of 8.4 mm diameter and 300-1000 micrometer were obtained. For half of the samples this deposition was performed without alignment field and for the other half an alignment field strength of 500 mT was used. 

\textit{Small-angle neutron scattering (SANS)}:
Neutron scattering experiments were performed at Budapest Neutron Centre (BNC) using neutron wavelengths of 4.2 and 9.7 nm, detector distances of 1.15 and 5.26 meters, and 10 mm beamstop \cite{alm2021a}. Disk samples were mounted with their diameter perpendicular to the beam, such that alignment field direction was perpendicular to the beam, with alignment field direction being oriented vertically. Scattering data was reduced and radially integrated according to \cite{alm2021a} using BerSANS. Normalization and background subtraction was performed in MATLAB. Fitting of model functions to the SANS data was performed in SasView 6.1. Fitting models are shown in the supplementary information.

\textit{Small-angle X-ray scattering (SAXS)}:
Small angle x-ray scattering was performed at the CoSAXS beamline at the MAXIV Laboratory in Lund, Sweden. 12.4 keV monochromatic beam was used with a beam size at the sample of roughly 100x100 $\mu$m \cite{MaxIV_CoSaxs}. Disk samples were mounted with their diameter perpendicular to the beam, such that alignment field direction was perpendicular to the beam, with the alignment field direction being oriented vertically. Several images were taken for each sample, scanning the sample for inhomogeneity. Data reduction, radial integration, and background subtraction performed in Python, using the MAXIV Laboratory computing infrastructure \cite{MaxIV_CoSaxs}. THe fitting of model functions to the SAXS data was performed in SasView 6.1. For information on the fitting models see supplementary information.
\begin{table}[bp!]
\centering
 \caption{Particle volume fraction, alignment field, sample weight, size (disk diameter and height), and demagnetisation factor along the diameter N$_{\textrm{d}}$ for samples used for VSM measurements.}
 \label{tab:VSMDemag}
 \begin{ruledtabular}
  \begin{tabular}{cc|ccc}
    \textrm{NP Cont. [vol\%]} & $H_{\textrm{align.}}$ & Mass [mg] & \textrm{Size (d$\times$h) [mm]} & \textrm{N$_{\textrm{d}}$}\\
    \colrule
    3.0 & 0 mT      &   204.5     & 12.0$\times$1.6      & 0.09 \\
    2.7 & 50 mT      & 213.5     & 11.7$\times$1.8      & 0.10\\
    2.8 & 100 mT      & 231.9     & 11.7$\times$1.9      & 0.11\\
    2.8 & 200 mT      & 211.3     & 11.2$\times$1.9      & 0.11\\
    2.7 & 500 mT      & 209.2     & 11.6$\times$1.8      & 0.10\\
    \hline
    12.4 & 0 mT      & 82.5     & 8.4$\times$1.00      & 0.14\\
    13.4 & 500 mT      & 75.0     & 8.4$\times$0.89      & 0.074\\
    30.5 & 0 mT      & 52.8     & 8.4$\times$0.44      & 0.046\\
    32.1 & 500 mT      & 49.7     & 8.4$\times$0.40      & 0.036\\
    56.1 & 0 mT      & 46.1     & 8.4$\times$0.26      & 0.024\\
    56.8 & 500 mT      & 46.4     & 8.4$\times$0.26      & 0.024
  \end{tabular}  
\end{ruledtabular}
\end{table}

\textit{Vibrating sample magnetometry (VSM)}:
DC-magnetisation measurements were performed on a Lake Shore Cryotronics 7407-series VSM. Measurements were made at room temperature with applied field up to 1 T, using the point-by-point mode, step size of 795.8 A/m (10 Oe), 0.1 second hardware time constant and 2 seconds software averaging time.
Disk samples were oriented with diameter parallel to the applied field, using the demagnetisation factors specified in \textbf{table \ref{tab:VSMDemag}}, which were calculated based on \cite{beleggia2005a,BahlDemagCyl}. Samples synthesised with alignment field were measured twice, once with alignment field direction parallel to the measuring field direction and once with the measuring field direction perpendicular to the alignment field direction.

\textit{AC-susceptibility}:
Susceptibility measurements between 10 Hz and 500 kHz were performed on a DynoMag AC-susceptometry system from RISE Research Institutes of Sweden, using 0.5 mT applied field. Disk samples were oriented with diameter parallel to the applied field. Samples for the AC-susceptibility and looptracer measurements were cut from the larger VSM Sample disks due to size constraints. Demagnetisation factors used in the AC-susceptibility and looptracer measurements can be seen in \textbf{table \ref{tab:AC_Demag}} and are calculated based on \cite{beleggia2005a,BahlDemagCyl} . AC-demagnetisation correction was performed according to \cite{goldfarb1991a}. Samples synthesised with alignment field were measured twice, once with alignment field direction parallel to the measuring field direction and once with the measuring field direction perpendicular to the alignment field direction.

\begin{table}
\centering
 \caption{Particle fraction, alignment field and measurement orientation, sample weight, sample size (disk diameter and height), and demagnetisation factor along the diameter N$_{\textrm{d}}$ (or height N$_{\textrm{h}}$) for samples used for AC-susceptometry and looptracer measurements.}
 \label{tab:AC_Demag}
 \begin{ruledtabular}
  \begin{tabular}{cc|ccc}
    \textrm{NP Cont.} & $H_{\textrm{align}}$ & Mass & \textrm{Size (d$\times$h)} & \textrm{N$_{\textrm{d}}$}\\
    \textrm{[vol\%]} & (dir) & [mg] & [mm] &  \\
    \colrule
    3 & 0 mT      &   49.3     & 4.9$\times$1.9      & 0.22 \\
    2.7 & 50 mT ($||$)     & 45.6     & 4.9$\times$1.8      & 0.21\\
    2.7 & 50 mT ($\perp$)     & 45.7     & 4.9$\times$1.8      & 0.21\\
    2.8 & 100 mT ($||$)     & 50.0     & 4.9$\times$2.0      & 0.22\\
    2.8 & 100 mT ($\perp$)     & 55     & 4.9$\times$2.2      & 0.23\\
    2.8 & 200 mT ($||$)     & 49.8     & 4.9$\times$1.9      & 0.22\\
    2.8 & 200 mT ($\perp$)      & 47.4     & 4.9$\times$1.9      & 0.22\\
    2.7 & 500 mT ($||$)     & 46.1     & 4.9$\times$1.8      & 0.21\\
    2.7 & 500 mT ($\perp$)     & 53.9     & 4.9$\times$2.1      & 0.23\\
    \hline
    12.4 & 0 mT      & 82.5     & 8.4$\times$1.0      & 0.140\\
    13.4 & 500 mT ($||$)      & 75.0     & 8.4$\times$0.89      & 0.140\\
    13.4 & 500 mT ($\perp$)     & 75.0     & 8.4$\times$0.89      & 0.140\\
    30.5 & 0 mT      & 52.8     & 8.4$\times$0.44      & 0.065\\
    31.8 & 500 mT ($||$)     & 49.7     & 8.4$\times$0.40      & 0.044\\
    32.3 & 500 mT ($\perp$)     & 49.7     & 8.4$\times$0.40      & 0.044\\
    56.1 & 0 mT      &  20.0    & 6.2$\times$0.21      & 0.017\\
    56.5 & 500 mT ($||$)     & 11.3     & Ellipse: 0.71$\times$8.4      & N$_{\textrm{h}}$: 0.0145\\
    57.0 & 500 mT ($\perp$)     & 15.2     & Ellipse: 1.1$\times$6      &  0.073
  \end{tabular}  
\end{ruledtabular}
\end{table}

\textit{High-frequency hysteresis measurements}:
High-field dynamic magnetization measurements were done at 160.6, 241, 404, 570.6, and 922.7 kHz on a looptracer system using a MagneTherm as excitation source, two counter-would pick-up coils, and a Zürich Instruments MFLI 5 MHz lock-in amplifier \cite{TVeile2025}. Applied field was tuned to a $B$-field amplitude of 10 mT for each sample and frequency. The system was calibrated to magnetic moment and phase angle by use of a secondary coil excited by the lock-in amplifier tone generator, and by use of paramagnetic Dy$_2$O$_3$ powder sample for each frequency and applied field. Demagnetisation and measurement orientations are identical to the AC-susceptibility measurements.

\section*{Acknowledgments}
The authors thank the Independent Research Fund Denmark (project HiFMag, grant number 9041-00231A), the Danish Agency for Science, Technology, and Innovation (instrument center DanScatt, grant number 7129-00006B) and the Taumose Legat (grant 2022) for financial support.
The authors acknowledge the MAXIV Laboratory for beamtime on the CoSAXS beamline under proposal 20221115. Research conducted at MAX IV, a Swedish national user facility, is supported by Vetenskapsrådet (Swedish Research Council, VR) under contract 2018-07152, Vinnova (Swedish Governmental Agency for Innovation Systems) under contract 2018-04969 and Formas under contract 2019-02496.

\bibliography{Bibliography}

\end{document}